\documentclass{article}
\usepackage{graphicx}
\usepackage{wrapfig}
\usepackage{subfigure}
\usepackage{booktabs}
\usepackage{color}
\usepackage{cancel}
\usepackage{soul}

\title{\textbf{\large{Kelvin-Helmholtz and Buckling Instabilities for a Viscoelastic Liquid}}}
\author{\textsc{\small{Bavand Keshavarz$^*$, and Gareth H. McKinley}}\\ \small{\textit{Department of Mechanical Engineering, Massachusetts Institute of Technology}}\\ \small{\textit{77 Massachusetts Avenue, Cambridge - MA 02139, USA}}\\ {\small $^*$Corresponding Author: bavand@mit.edu}}
\date{October, 14, 2011}

\begin{document}
\maketitle

\begin{abstract} 
In this fluid dynamics video prepared for the APS-DFD Gallery of Fluid Motion we study the Kelvin-Helmholtz instability for both Newtonian and viscoelastic jets. The nonlinear dynamics of the jet motion are slowed down by orders of magnitude using a synchronized strobe effect coupled with precise timing control of perturbation frequencies. Our results show that at high wave-numbers the imposed perturbations initially grow linearly with time and the jet axis remains straight while the Kelvin-Helmholtz wave amplitude grows and rolls up into bags that encapsulate the central jet within themselves.  At low wave-numbers (long wave-lengths) the jet axis buckles under the action of viscous stresses and a coupling between the Kelvin-Helmholtz instability and bending of the jet leads to new concertina or chevron modes which grow with time as they move downstream. Addition of viscoelasticity to the jet leads to the pronounced inhibition of the Kelvin-Helmholtz instability as the jet perturbation amplitude grows and large elongational stresses in the fluid become important.  For long waves, the initially-relaxed viscoelastic jet first buckles in a manner similar to the Newtonian solvent but once again the viscoelastic effects suppress the instability growth as they are convected downstream.  
\end{abstract} 

\section{\textbf{{ Introduction}}} \label{sec:introduction}
Kelvin-Helmholtz instability is a well-known phenomenon, which occurs in nature when two identical or dissimilar liquids with different velocities meet each other at an interface. Wave generation on water surfaces in a windy day, shear layer patterns in clouds, and even wave patterns observed in solar winds represent common examples showing this instability in nature [1]. Kelvin-Helmholtz instability also occurs in the industrially-important process of air-assisted atomization (between two coaxial jets) as the primary instability generating waves on the liquid surface [3].
Helmholtz (1868) was the first scientist who noted this instability and remarked: Òevery perfect geometrically sharp edge by which a fluid flows must tear it asunder and establish a surface of separation, however slowly the rest of the fluid may moveÓ [1]. Lord Kelvin (1871) later developed this idea furthermore and established a mathematical solution for the problem. He later used the results to understand wave generation by wind blowing across a flat ocean surface. 
The effects of viscoelasticity on the Kelvin-Helmholtz instability are still largely unknown, despite the fact that viscoelastic jets and interfaces are key components of many industrial applications such as air-assisted atomization. The current study investigates the effects of viscoelasticity on this instability, in an axisymmetric flow configuration, benefitting from a novel strobe-based high Ðresolution experimental setup.

\section{\textbf{{ Experimental Setup}}} \label{sec:Experimental Setup}

The setup used in these experiments (Figure 1) has a sub-millimeter nozzle (diameter:$150 \mu m$ ) from which fluid is discharged into a water tank. The tank is made of transparent flat walls, which provide optical access to the axial position where instability starts. The nozzle is perturbed using an annular piezoelectric actuator squeezing the nozzle walls within a wide range of frequencies (${1Hz <f<40 kHz}$ ). The nozzle and the tank are both positioned on an two-axis positioning stage, mounted on a stepper motor for precise motion in the axial direction (Figure 1). Liquids are pumped using a syringe pump (4400 Ultra PHD from Harvard Apparatus) at a fixed rate with precise control on the pumping rate (errors in flow rate are less than 0.5 percent). Two signal generators are used for perturbing the jet; one carries the signal to the piezoelectric actuator at moderately small amplitudes (1-5 volts) and the other generator sends a similar signal with slightly different frequency  ${f-\delta f}$ to the strobe light. This slight difference in frequencies between the perturbation and the pulsed lighting leads to an apparent slowing down of the motion by a factor of ${\delta f / f}$ .
\begin{figure}
\begin{center}
\includegraphics[width=0.9\textwidth]{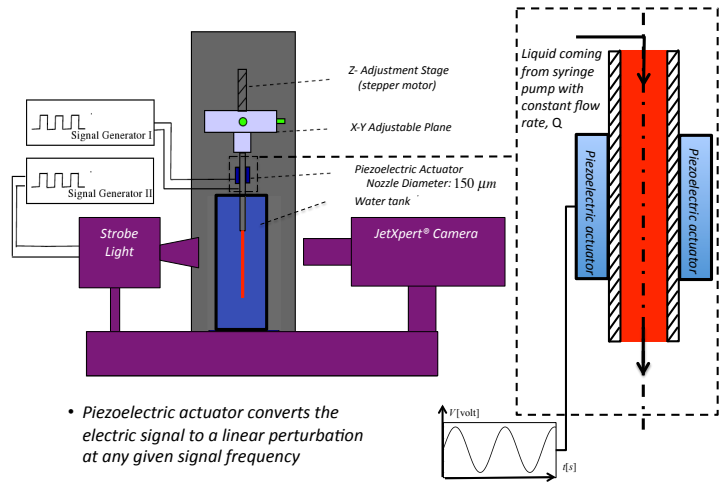}
\caption{The experimental set-up used in this study.}
\end{center}
\end{figure}

\section{\textbf{{ Test Fluids}}} \label{sec:Test Fluids}

The background liquid inside the tank is DI water, which is kept at room temperature. Two different liquids are discharged in the water tank; a Newtonian solvent (water and glycerol mixture) and a dilute viscoelastic solution made by dissolving a small amount (c = 0.1percent) of Poly-Ethylene-Oxide (PEO-300K) into the same Newtonian solvent. The shear viscosities of these two liquids are very close to each other at high shear rates (a common property of dilute PEO solutions) also their liquid/air surface tensions are almost identical since PEO is not active at the surface. The only difference between these two is the addition of extensional viscoelasticity, which results from dissolving PEO polymer chains in the solvent [4]. The material parameter that quantifies this difference is the extensional relaxation time (${\lambda_E}$ ).  For the Newtonian solvent tested (like any other Newtonian liquid) the relaxation time is zero since there is no intermolecular mechanism for elasticity or stress relaxation, however by addition of 0.1 percent of PEO the relaxation time increases to ${\lambda_E=0.36 \, ms}$(measurements were done using jet break-up extensional rheometry [5]).

\section{\textbf{{ Contents of the Video}}} \label{sec:Contents of the Video }

As the Newtonian liquid leaves the nozzle and discharges into the water tank with a velocity around 3 m/s, shear instabilities start to appear due to the difference in velocities at the liquid-liquid interface. The amplitude of the wave crests grow with time as the jet moves downstream and become elongated in the upstream direction as they protrude into the stationary outer fluid; then due to the axisymmetry of the geometry, the wave crests form re-entrant ÒbagsÓ that wrap around the jet axis entraining the outer fluid. The captured movies show that at high wave-numbers (or short wavelengths) the Kelvin-Helmholtz instability develops for the Newtonian solvent right after the nozzle exit and the instability evolves rapidly into a nonlinear state consisting of stacked chevron-like waves. At these high wave-numbers the jet remains axially straight throughout the evolution of the Kelvin-Helmholtz instability and there is no sign of bending or buckling in the jet axis (Figure 2). At low wave-numbers (long wavelengths) for the Newtonian liquid the shear instabilities start again close to the nozzle exit but at the same time the jetÕs axis starts to buckle due to the viscous compressive stresses acting on the fluid column [6]. Bending of the jet, coupled with the nonlinear evolution of the waves resulting from Kelvin-Helmholtz instability, leads to a complex sinuous wave pattern of growing amplitude (which has the same frequency as the imposed perturbation). The buckling and associated deceleration in the viscous jet concertinas each wave crest into the next one in a shape very similar to what one observes in tulip petals. These Òtulip likeÓ patterns are convected downstream and grow approximately linearly in radius with time while becoming progressively compressed and stacked together in the z-direction (Figure 3).

\begin{figure}
\begin{center}
\includegraphics[width=1\textwidth]{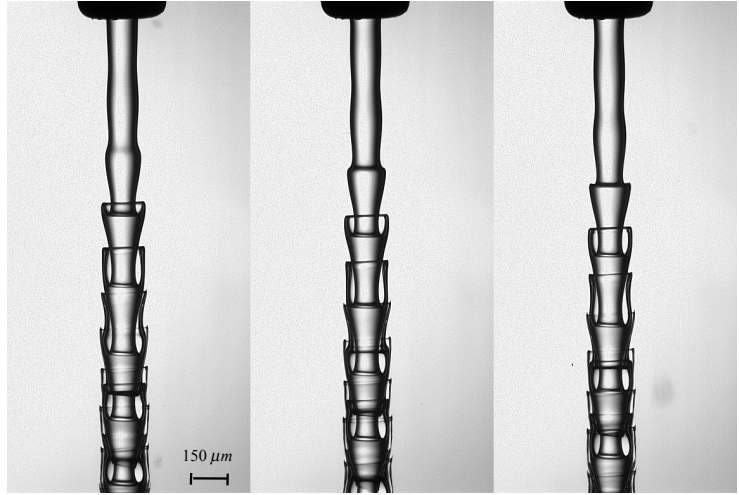}
\caption{\small Three snapshots of the Kelvin-Helmholtz instability (with $40 \mu s$  time difference) for the Newtonian liquid and short perturbation wavelengths.}
\small${kR_0 \equiv \left ( 2\pi f/V \right )R_0 =0.8\,,\,\rho/\rho_w=1.103\,,\,\mu/\mu_w=3.2\,,\,}$
\small \\${Re\equiv \left(\rho VD/\mu \right )=146\,,\,Oh\equiv \mu/\sqrt{\rho\,\sigma D}=0.1\,,\,Wi\equiv \lambda_E V/D=0}$
\end{center}
\end{figure}

\begin{figure}
\begin{center}
\includegraphics[width=1\textwidth]{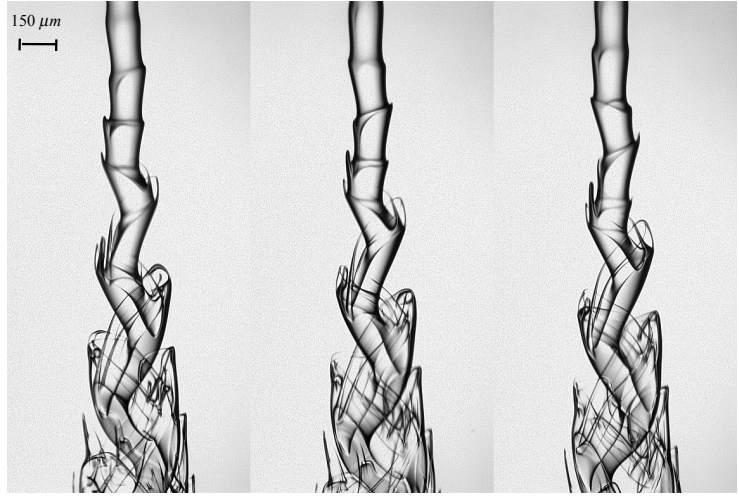}
\caption{\small Three snapshots of the Kelvin-Helmholtz instability (with $40 \mu s$  time difference) for the Newtonian liquid and long perturbation wavelengths.}
\small${kR_0 \equiv \left ( 2\pi f/V \right )R_0 =0.3\,,\,\rho/\rho_w=1.103\,,\,\mu/\mu_w=3.2\,,\,}$
\small \\${Re\equiv \left(\rho VD/\mu \right )=146\,,\,Oh\equiv \mu/\sqrt{\rho\,\sigma D}=0.1\,,\,Wi\equiv \lambda_E V/D=0}$
\end{center}
\end{figure}

For the viscoelastic jet, at high wave-numbers (short wavelengths), the Kelvin-Helmholtz instability again starts from the nozzle and in the linear region the instability behaves very similarly to the Newtonian solvent. However, as the instability grows to larger amplitudes the wave crests show enhanced resistance to the elongation due to the presence of viscoelasticity and this consequently inhibits the nonlinear bag formation and encapsulation process (Figure 4). At short wave-numbers (long wavelengths) the viscoelastic jet again undergoes a buckling instability coupled with Kelvin-Helmholtz perturbations very similar to the Newtonian jet (Figure 5) but the tulip patterns are not convected downstream in a regular way and the width of the jet grows much more slowly than was observed in the Newtonian solvent due to the additional viscoelastic stresses.

\begin{figure}
\begin{center}
\includegraphics[width=1\textwidth]{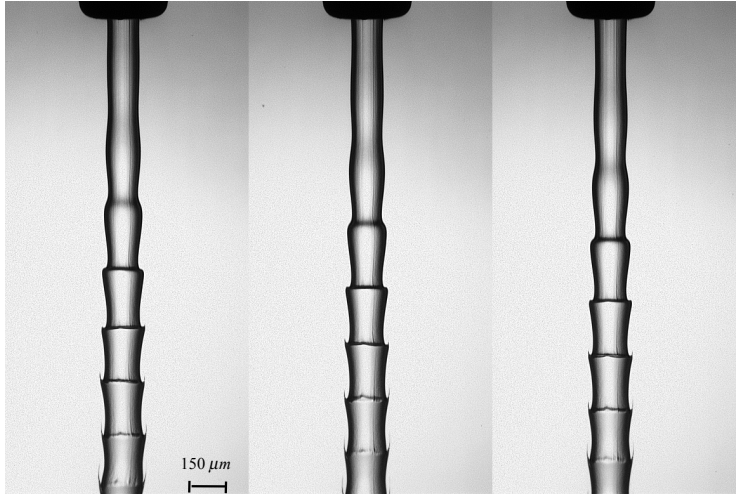}
\caption{\small Three snapshots of the short-wavelengthKelvin-Helmholtz instability (with $40 \mu s$  time difference) for the viscoealstic liquid.}
\small${kR_0 \equiv \left ( 2\pi f/V \right )R_0 =0.8\,,\,\rho/\rho_w=1.103\,,\,\mu/\mu_w=3.2\,,\,}$
\small \\${Re\equiv \left(\rho VD/\mu \right )=146\,,\,Oh\equiv \mu/\sqrt{\rho\,\sigma D}=0.1\,,\,Wi\equiv \lambda_E V/D=0}$
\end{center}
\end{figure}

\begin{figure}
\begin{center}
\includegraphics[width=1\textwidth]{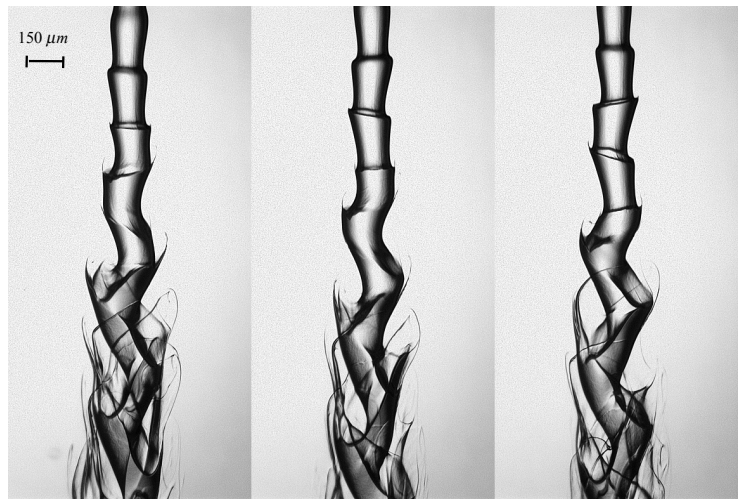}
\caption{\small Three snapshots of the Kelvin-Helmholtz instability (with $40 \mu s$  time difference) for the viscoelastic liquid and long perturbation wavelengths.}
\small${kR_0 \equiv \left ( 2\pi f/V \right )R_0 =0.3\,,\,\rho/\rho_w=1.103\,,\,\mu/\mu_w=3.2\,,\,}$
\small \\${Re\equiv \left(\rho VD/\mu \right )=146\,,\,Oh\equiv \mu/\sqrt{\rho\,\sigma D}=0.1\,,\,Wi\equiv \lambda_E V/D=0}$
\end{center}
\end{figure}
\newpage
\section{\textbf{{ References}}} \label{sec:References}

\begin{enumerate}
\item{ Drazin P.G., Introduction to Hydrodynamic Instability, Cambridge Texts in Applied Mathematics, 2002.}
\item{ Chandrasekhar S., Hydrodynamic and Hydromagnetic Instability, Dover Publication, 1981.}
\item{ Marmottant P., and Villermaux, E., On Spray Formation, Journal of Fluid Mechanics, 2004, 498:73-111.}
\item{ Rodd L.E., Scott T.P., Cooper-White J.J., and McKinley G.H., Capillary Break-up Rheometry of Low-Viscosity Elastic Fluids, Applied Rheology, 2005, 15:12-27. }
\item{ Ardekani, A.M., Sharma, V., and McKinley G.H., Dynamics of Bead Formation, Filament Thinning, and Break-up in Weakly Viscoelastic Jets, Journal of Fluid Mechanics, 2010, 665:46-56. }
\item{ Taylor G.I., Instability of Jets Threads and Sheets of Viscous Fluid, Proceedings of the 12th International Congress in Applied Mechanics, Stanford, 1968, 382-388. }
\end{enumerate}

\end{document}